\documentclass[showpacs,prl,onecolumn,aps,superscriptaddress,preprintnumbers,letterpaper]{revtex4}
\usepackage{amsmath,amssymb}
\usepackage{epsfig}
\usepackage{graphicx}
\usepackage{amsmath}
\usepackage{amsfonts}
\usepackage{epstopdf}
\def\slashchar#1{\setbox0=\hbox{$#1$}     		
   \dimen0=\wd0                                 	
   \setbox1=\hbox{/} \dimen1=\wd1               	
   \ifdim\dimen0>\dimen1                        	
      \rlap{\hbox to \dimen0{\hfil/\hfil}}      	
      #1                                        	
   \else                                        	
      \rlap{\hbox to \dimen1{\hfil$#1$\hfil}}   	
      /                                         	
   \fi}

\newcommand{\be}{\begin{equation}}
\newcommand{\ee}{\end{equation}}
\newcommand{\bear}{\begin{eqnarray}}
\newcommand{\eear}{\end{eqnarray}}
\newcommand{\ba}{\begin{array}}
\newcommand{\ea}{\end{array}}

\newcommand{\avg}[1]{\left \langle #1 \right \rangle}
\begin{document}

\title{Jet energy loss and fragmentation in heavy ion collisions }

\author{Dmitri E. Kharzeev}
\affiliation{Department of Physics and Astronomy, Stony Brook University, Stony Brook, New York 11794-3800, USA}
\affiliation{Department of Physics,
Brookhaven National Laboratory, Upton, New York 11973-5000, USA}

\author{Frash\"er Loshaj}
\affiliation{Department of Physics and Astronomy, Stony Brook University, Stony Brook, New York 11794-3800, USA}

\date{\today}

\begin{abstract}
Recent LHC results indicate a suppression of jet fragmentation functions in Pb-Pb collisions at intermediate values of $\xi=\ln(1/z)$.
 This seems to contradict the picture of energy loss based on the induced QCD radiation that is expected to lead to the enhancement of in-medium fragmentation functions. We use an effective $1+1$ dimensional quasi-Abelian model to describe the dynamical modification of jet fragmentation in the medium. We find that this approach describes the data, and argue that there is no contradiction between the LHC results and the picture of QCD radiation induced by the in-medium scattering of the jet.  The physics that underlies the suppression of the in-medium fragmentation function at intermediate values of $\xi=\ln(1/z)$ is the partial screening of the color charge of the jet by the comoving medium-induced gluon. 
 
 \end{abstract}

\maketitle

Recently, the CMS Collaboration presented the data on the modification of the shape of the jets produced in Pb-Pb collisions at the LHC \cite{CMS:2012wxa}. 
This data is interesting because it opens a window into the mechanism by which jets lose energy in the quark-gluon plasma (QGP). It is expected that the dominant mechanism of 
jet energy loss is the induced QCD bremsstrahlung \cite{Gyulassy:1993hr,Baier:1996kr,Gyulassy:2000er,CasalderreySolana:2010eh,Zakharov:1996fv,Baier:2000mf}. This induced gluon radiation would then transform into hadrons and produce 
an enhancement in the in-medium jet fragmentation function at small values of $z$, the fraction of the jet's energy carried by the produced hadron, or equivalently, at large values of $\xi = \ln (1/z)$ -- see \cite{Armesto:2011ht} for a recent overview and comparison of various models of energy loss. The data indeed clearly show this enhancement \cite{CMS:2012wxa}. However, the data also indicate the {\it suppression} of the fragmentation function at intermediate values of $\xi \simeq 3$.  This suppression is surprising because it seems to imply, through the Local Parton Hadron Duality (LPHD) \cite{Dokshitzer:1991wu}, that the radiation of gluons at these intermediate values of $\xi$ is suppressed relative to the in-vacuum fragmentation. This apparent suppression of gluon radiation is hard to reconcile with the expected enhancement due to the induced  QCD bremsstrahlung. In this paper we use an effective model of jet fragmentation \cite{Loshaj:2011jx} to argue that there is no contradiction between the CMS result and the presence of induced QCD radiation. 
\vskip0.3cm

Describing the fragmentation of a jet into hadrons from first principles requires a theory of confinement, and it is still lacking. Instead, the conventional pQCD approach \cite{Collins:1989gx} is based on introducing universal phenomenological fragmentation functions extracted from the experimental data. 
This practical and useful approach however does not allow to predict how these fragmentation functions would change in the presence of the QCD medium -- making such a prediction requires a dynamical theory of fragmentation. While the complete theory of confinement still does not exist, many properties of 
confining interactions in QCD are known from phenomenology, lattice QCD, and effective theories. One of the properties of QCD with light quarks is the so-called  ``soft confinement" \cite{Gribov:1999ui} (for review, see e.g. \cite{Dokshitzer:2004ie}). For the case of jet fragmentation, ``soft confinement" implies that 
the fragmenting quark polarizes the QCD vacuum and slows down by producing along its trajectory quark-antiquark pairs that later form hadrons -- this picture in fact can be considered as the foundation of the phenomenologically successful LPHD hypothesis. 
\vskip0.3cm

The massless QED in $1+1$ dimensions (${\rm QED}_2$, also known as the Schwinger model \cite{Schwinger:1962tp,Lowenstein:1971fc,Coleman:1975pw}) was proposed long time ago 
as an effective theory of quark fragmentation in $e^+e^-$ annihilation \cite{:1974cks}. Indeed, this exactly soluble model captures many properties of quark interactions in QCD -- the screening of color charge by light quark-antiquark pairs, the presence of $\theta$-vacuum, and the axial anomaly. QED$_2$ has previously been applied to the description of hadronic interactions at high energies in Refs. \cite{Fujita:1989vv,Wong:1991ub}.

The use of the Abelian ${\rm QED}_2$ model for describing jet fragmentation could be justified by i) the effective dimensional reduction that occurs for high momentum quarks and ii) the picture of confinement based on the condensation of magnetic monopoles in QCD vacuum and the resulting quasi-Abelian projection \cite{'tHooft:1977hy,Mandelstam:1974pi}. In $3+1$ dimensions, the typical transverse momentum of mesons is of the order of their mass (in our case $m \simeq 600$ MeV), therefore their longitudinal momentum $p$ is much larger for $p E_{jet} = z>0.01$ or $\xi<5$ for $E_{jet}\sim 120$ GeV. A natural extension \cite{Loshaj:2011jx} of the model is thus to consider $N_c$ copies of the Abelian $U(1)$ gauge group. For the propagation of the quark jet through the medium, this extension allows to consider the rotation of the color orientation of the quark. By using the (1+1) dimensional field theory we neglect the transverse momentum broadening of the jet in the medium; this is a reasonable approximation for the
high momentum jets that we consider. Every time the quark exchanges a gluon with the medium, its color changes; for a medium of length $L$ and the quark mean free path $\lambda$, we thus get $L/\lambda$ sectors bounded by the propagating quark and the exchanged gluons, see Fig. \ref{fig:scatt_1_2}. At large $N_c$, these sectors produce particles  independently from each other. 
\vskip0.3cm

Let us recall the model used in \cite{Loshaj:2011jx}. We start from massless QED$_2$,    
\begin{equation}
\mathcal{L}=-\frac{1}{4}F_{\mu\nu}F^{\mu\nu}+\bar{\psi}(i\gamma^\mu\partial_\mu-g\gamma^\mu A_\mu)\psi
\label{eq:lang}
\end{equation}
The coupling constant $g$ in this theory has dimension of mass. In what follows, we will label the space-time coordinate by $x^\mu=(t,z)$. It is well known that in $1+1$ dimensions bosonization is an exact and a very convenient method, so we will rely on it. The bosonized form of the vector current is 
\begin{equation}
j^\mu(x)=\psi(x)\gamma^\mu\psi(x)=\frac{1}{\sqrt{\pi}}\epsilon^{\mu\nu}\partial_\nu\phi(x)
\label{eq:bos}
\end{equation}
where $\phi(x)$ is the Klein-Gordon scalar field. We introduce the jet into the system by coupling an external $U(1)$ current $j^\mu_{\mathrm{ext}}(x)$ to the gauge field $A_\mu$. We construct this current in a usual way from the classical trajectory $y(\tau)$ of the $U(1)$ charge
\begin{equation}
j_{\mathrm{ext}}^\mu(x)=\int{d\tau\frac{dy^\mu(\tau)}{d\tau}\delta^{(2)}(x-y(\tau))} .
\label{eq:trcurr}
\end{equation}
After adding the term $j_{\mathrm{ext}}^\mu(x) A_\mu(x)$ in \eqref{eq:lang}, and integrating out the gauge field, we get the effective lagrangian
\begin{equation}
\mathcal{L}=\frac{1}{2}(\partial_\mu\phi)^2-\frac{1}{2}\frac{g^2}{\pi}(\phi+\phi_{\mathrm{ext}})^2 ,
\label{eq:efflang}
\end{equation}
where we have used the parametrization
\begin{equation}
j_{\mathrm{ext}}^\mu(x)=\frac{1}{\sqrt{\pi}}\epsilon^{\mu\nu}\partial_\nu\phi_{\mathrm{ext}}(x) .
\label{eq:currpar}
\end{equation}
From \eqref{eq:efflang} we get the equation of motion
\begin{equation}
(\Box+m^2)\phi(x)=-m^2\phi_{\mathrm{ext}}(x) ,
\label{eq:eom}
\end{equation}
where $m =  g/\sqrt{\pi}$. 
\vskip0.3cm

\begin{figure}[htbp]
\centering
\includegraphics[width=.45\linewidth]{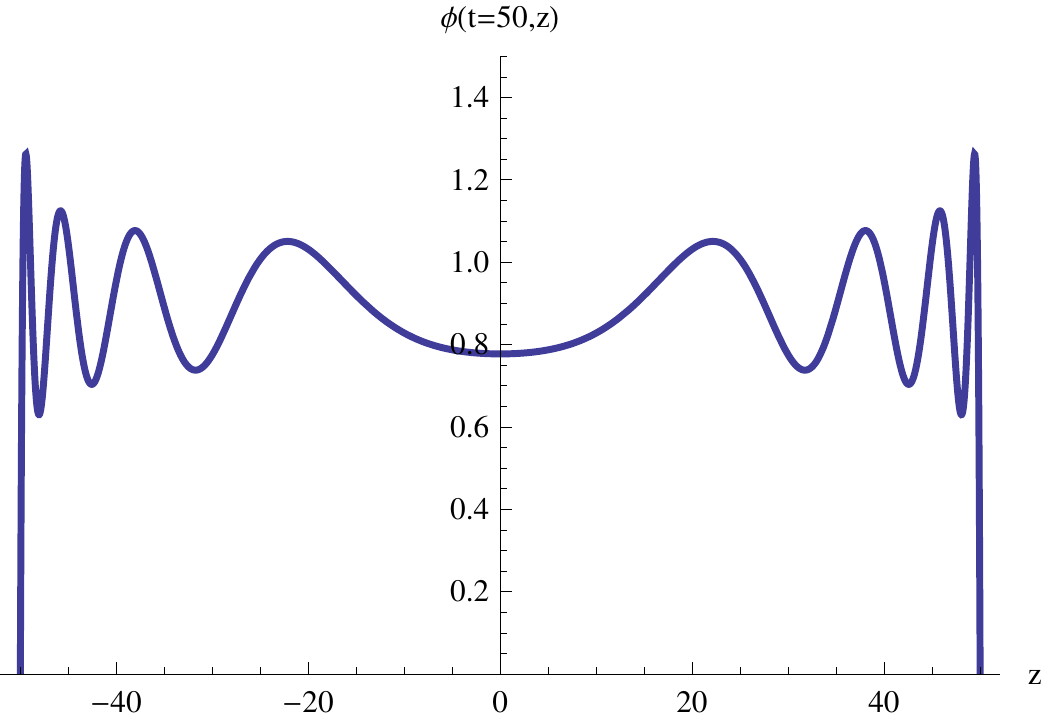} 
\caption{Scalar field $\phi$ as a function of the spatial coordinate $z$ for $m=0.6 \ \mathrm{GeV}$  and $t=10 \ \mathrm{fm}$.}
\label{fig:sbreak}
\end{figure}

Let us consider an external source consisting of a charge and an anti-charge moving back-to-back along the light cone, namely
\begin{equation}
j^0_{\mathrm{ext}}(x)=-\delta(z+t)\theta(-z)+\delta(z-t)\theta(z) .
\label{eq:lcs}
\end{equation}
With this source, we can solve \eqref{eq:eom} exactly:
\begin{equation}
\phi(x)=\theta(t^2-z^2)[1-J_0(m \sqrt{t^2-z^2})]
\label{eq:sol}
\end{equation}
The result is plotted in Fig. \ref{fig:sbreak} for fixed values of $m$ and $t$.
The interpretation of Fig.\ref{fig:sbreak} is as follows. Since in $1+1$ dimensions the potential between charge and anti-charge at short distances is linear, initially we form a string. We can see that the string formed between receding particles breaks into the quark-antiquark pairs  -- indeed, as it follows from \eqref{eq:bos}, the kinks and anti-kinks of the scalar field represent the charged fermions and anti-fermions respectively. 
The momentum distribution of particles produced by the classical source $-m^2\phi_{\mathrm{ext}}(x)$ coupled to the scalar field can be written as
\begin{equation}
\frac{dN}{dp}=\frac{1}{2\omega}|-m^2\tilde{\phi}_{\mathrm{ext}}(p)|^2 ,
\label{eq:spec}
\end{equation}
where $\tilde{\phi}_{\mathrm{ext}}(p)$ is the Fourier transform of $\phi_{\mathrm{ext}}(x)$, $p^\mu=(\omega, \ p)$ and $\omega=\sqrt{p^2+m^2}$. 
\vskip0.3cm

The prescription of using this model is thus to first use \eqref{eq:trcurr} to construct the Abelian current and then use the parametrization \eqref{eq:currpar} to get $\phi_{\mathrm{ext}}(x)$. An application of the model to $e^+e^-$ annihilation as well as to the in-medium scattering was already presented in \cite{Loshaj:2011jx}. In the latter case, we assumed the static scattering centers with no momentum transfer between the medium and the jet. The main deficiency of our approach in \cite{Loshaj:2011jx} was the absence of the perturbative medium-induced gluon radiation. In the conventional pQCD picture, it is this medium-induced radiation modified severely \cite{Baier:1996kr} by the Landau-Pomeranchuk-Migdal (LPM) effect that is responsible for the jet energy loss. In our approach in \cite{Loshaj:2011jx} the energy loss was entirely due to the enhanced soft hadron production induced by the presence of additional color sectors due to the rotation of jet color polarization in the medium. 
\vskip0.3cm

In our current treatment we improve on the approach of \cite{Loshaj:2011jx} by considering the medium-induced perturbative gluon radiation. Compared to the conventional pQCD approaches, we consider also the dynamical modification of the in-medium jet fragmentation due to both the multiple scattering of the jet in the medium and the induced gluon radiation. In the present paper we also allow for a non-zero momentum transfer from the jet to the medium.
 Due to this momentum transfer, the scattered particles in the medium are given a kick along the jet momentum and move with some finite velocity after scattering. The typical momentum transfer in medium is of the order of Debye mass $m_D$, which is also a typical mass of the scattered particles. We can therefore estimate the velocity to be $v_i\sim 1/\sqrt{2}$ ($i=1,\cdots,n$). Let us assume first that the induced radiation is emitted outside of the medium -- this process is illustrated in Fig. \ref{fig:scatt_1_2};  the quark jet scatters $n$ times within the medium prior to emitting a gluon, while the corresponding antiquark jet is assumed to escape without interactions, as would be the case for the surface emission. On the right in Fig. \ref{fig:scatt_1_2} we show the corresponding color flow; each color contour at large $N_c$ radiates independently, as explained above. The trajectory of the Abelian charge is given by the boundary of the contour. There are only three different types of currents we have to consider and they are labeled by $j_1$, $j_2$ and $j_3$. Using the prescription above we can write down the charge densities:
\begin{figure}[htbp]
\centering
\includegraphics[width=.8\linewidth]{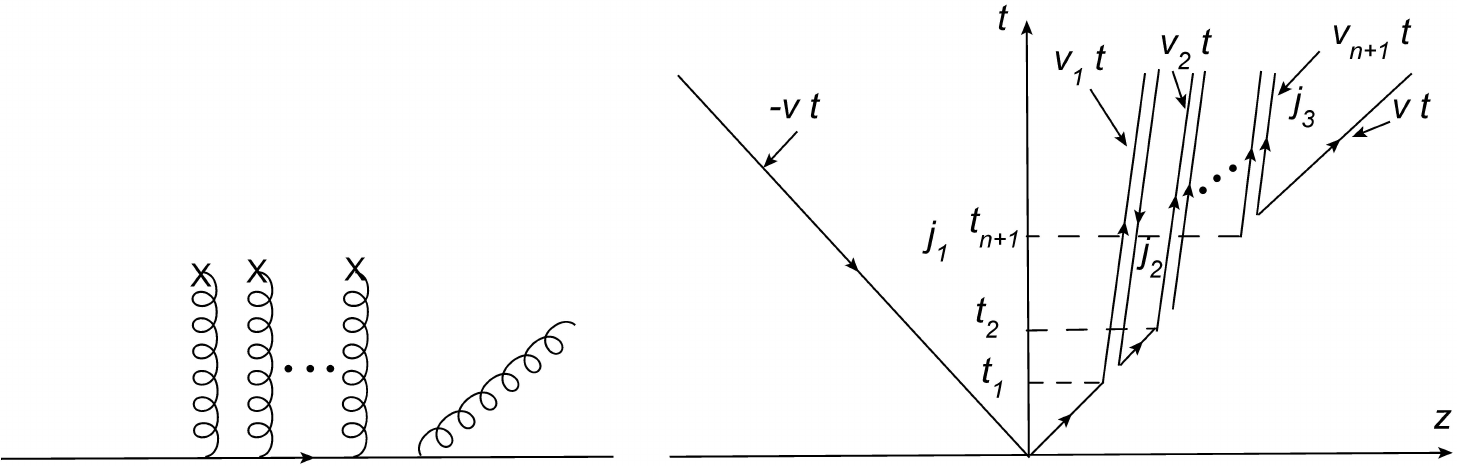} 
\caption{In-medium scattering of the jet accompanied by an induced gluon radiation outside of the medium. Left: the Feynman diagram. Right: the corresponding color flow.}
\label{fig:scatt_1_2}
\end{figure}
\begin{eqnarray}
j_1^0(x)&=&-\delta(z+vt)\theta(-z)+\big\{\delta(z+vt)[\theta(t)-\theta(t-t_1)] \nonumber \\ 
&+& \delta[z-vt_1-v_1(t-t_1)]\theta(t-t_1)\big\}\theta(z) \nonumber \\
j_2^0(x)&=&-\delta[z-vt_1-v_1(t-t_1)]\theta(t-t_1) + \delta(z-vt)[\theta(t-t_1)-\theta(t-t_2)] \nonumber \\ 
&+& \delta[z-vt_2-v_2(t-t_2)]\theta(t-t_2) \nonumber \\
j_3^0(x)&=&\big[-\delta[z-vt_{n+1}-v_{n+1}(t-t_{n+1})]\theta(t-t_{n+1}) +\delta(z-vt)\big]\theta(t-t_{n+1})
\label{eq:curr}
\end{eqnarray}
The Fourier transform of these charge densities is given by
\begin{eqnarray}
\tilde{j}_1^0(p)=\frac{ip}{\omega-vp}\Big[\frac{2v}{\omega+vp} -\frac{v-v_1}{\omega-v_1 p}e^{i(\omega-vp)t_1}\Big]
\label{eq:j1}
\end{eqnarray}
 
\begin{eqnarray}
\tilde{j}_2^0(p)=\frac{-ip}{\omega-vp}\Big[\frac{v-v_2}{\omega-v_2p}e^{i(\omega-vp)t_2} -\frac{v-v_1}{\omega-v_1p}e^{i(\omega-vp)t_1}\Big]
\label{eq:j2}
\end{eqnarray}
\begin{eqnarray}
\tilde{j}_3^0(p)=\frac{ip}{\omega-vp}\frac{v-v_{n+1}}{\omega-v_{n+1}p}e^{i(\omega-vp)t_{n+1}}
\label{eq:j3}
\end{eqnarray}
We can now use \eqref{eq:currpar} to construct the corresponding $\tilde{\phi}_{\mathrm{ext}}(p)$ to compute the distribution \eqref{eq:spec}. Please note that we only need $j^0_{\mathrm{ext}}$ to construct $\phi_{\mathrm{ext}}$. We can also get $j^1_{\mathrm{ext}}$ from $j^0_{\mathrm{ext}}$ by using the fact that the vector current is conserved. Since there is no interference between contours, we can write
\begin{equation}
\frac{dN^{med}}{dp}=\frac{m^4}{2 \omega}(|\tilde{\phi}_{1,\mathrm{ext}}(p)|^2+\sum{|\tilde{\phi}_{2,\mathrm{ext}}(p)|^2}+|\tilde{\phi}_{3,\mathrm{ext}}(p)|^2) ,
\label{eq:med}
\end{equation}
where $\tilde{\phi}_{2,\mathrm{ext}}(p)$ is calculated from $j_2^0(x)$ and we have to sum over all of contours of this type. We define $z= p/p_{\mathrm{jet}}$, where $p$ is the momentum of the final-state hadron and $p_{\mathrm{jet}}$ is the jet momentum. In order to compare with the data, we also define $\xi=\ln(1/z)$. 
\vskip0.3cm

\begin{figure}[htbp]
\centering
\includegraphics[width=.55\linewidth]{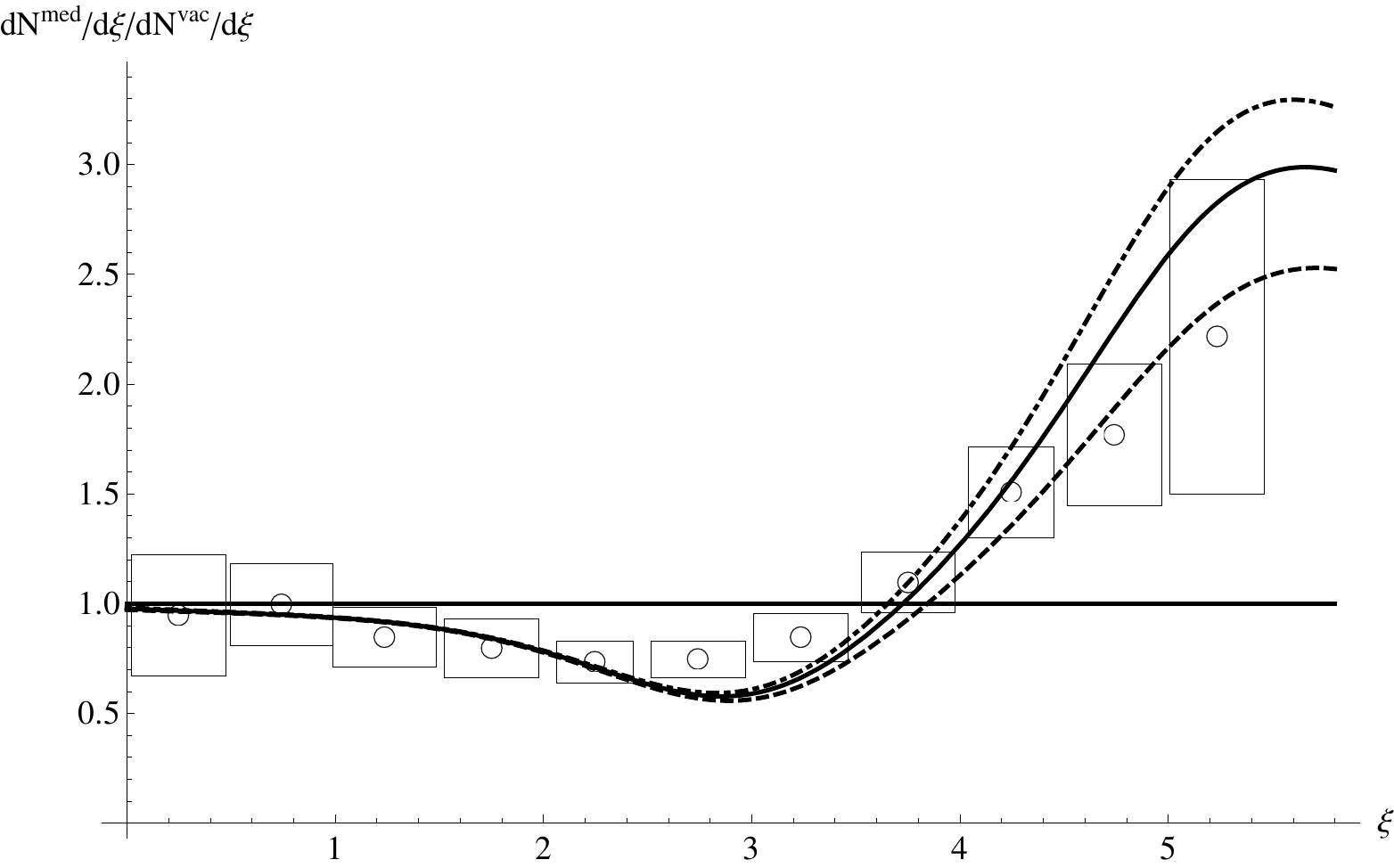} 
\caption{The ratio of in-medium and vacuum fragmentation functions for $p_{\mathrm{jet}}=120$ GeV. The first scattering occurs at $t_1\simeq 1$ fm, which is the assumed thermalization time. The length of the medium is $L = 5$ fm. The curves correspond to mean free paths of $\lambda = 0.57$, $0.4$ and $0.2$ fm from top to bottom respectively.}
\label{fig:frag_ratio}
\end{figure}

\begin{figure}[htbp]
\centering
\includegraphics[width=.8\linewidth]{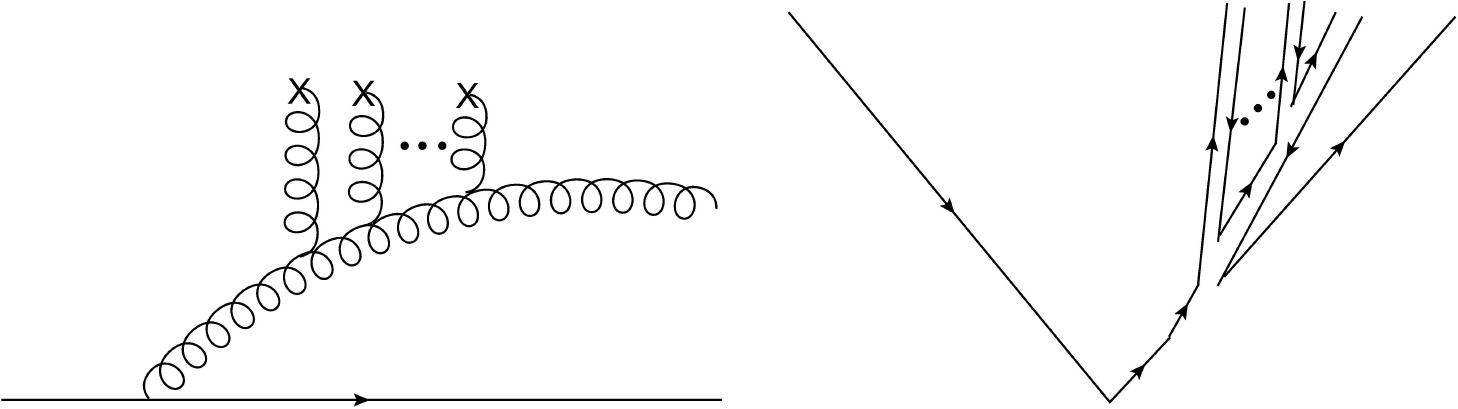} 
\caption{In-medium scattering of a gluon radiated from the original jet.}
\label{fig:scatt_2_2}
\end{figure}

We are now in a position to compute $dN^{\mathrm{med}}/d\xi$. Since we are interested in evaluating the ratio of the in-medium to in-vacuum fragmentation functions, we also need $dN^{\mathrm{vac}}/d\xi$ which has been evaluated already in \cite{Loshaj:2011jx} using as external source 
\begin{equation}
j_0(x)=-\delta(z-vt)\theta(-z)+\delta(z+vt)\theta(z) ,
\label{eq:curr_1}
\end{equation} 
where $v=p_{\mathrm{jet}}/\sqrt{p_{\mathrm{jet}}^2+Q_0^2}$ (same $v$ as in the expressions above) and $Q_0$ is in the range $1-3$ GeV. The ratio of fragmentation functions of in-medium and vacuum, as a function of $\xi=\ln\frac{1}{z}$, is shown in Fig. \ref{fig:frag_ratio}.
The result is plotted for different mean free paths, i.e. the different distances between scattering centers in Fig. \ref{fig:scatt_1_2}.  As mentioned above and as it was shown already in \cite{Loshaj:2011jx}, the enhancement for large $\xi$ results from the radiation coming from the medium-induced color  contours of type $j_2$. On the other hand, it can be seen that the suppression for intermediate $\xi$ comes from the contour of \eqref{eq:j3}. The underlying physics is the partial screening of the color charge of the jet by a comoving medium-induced gluon. A similar effect due to coherent parton branching has recently been considered in \cite{CasalderreySolana:2012ef}. When $v_{n+1}$ approaches $v$, where $v_{n+1}$ is the velocity of the final state gluon, we get a suppression in the fragmentation function. The final state gluon is typically emitted at the rapidity interval $\Delta \eta \sim 1/\alpha_{\rm s} \simeq 2$ away from the leading parton in the jet; this is the value that was assumed in the plot in Fig. \ref{fig:frag_ratio}.
 We have also considered the case when a gluon is radiated from the original jet and then interacts within the medium as shown in Fig. \ref{fig:scatt_2_2} -- this is the dominant diagram in the BDMPS \cite{Baier:1996kr,Gyulassy:2000er} approach. We have found that for the same values of parameters this case leads to the ratio of fragmentation functions  that is very similar to the one presented in Fig. \ref{fig:frag_ratio}.
\vskip0.3cm 

\begin{figure}[htbp]
\centering
\includegraphics[width=.55\linewidth]{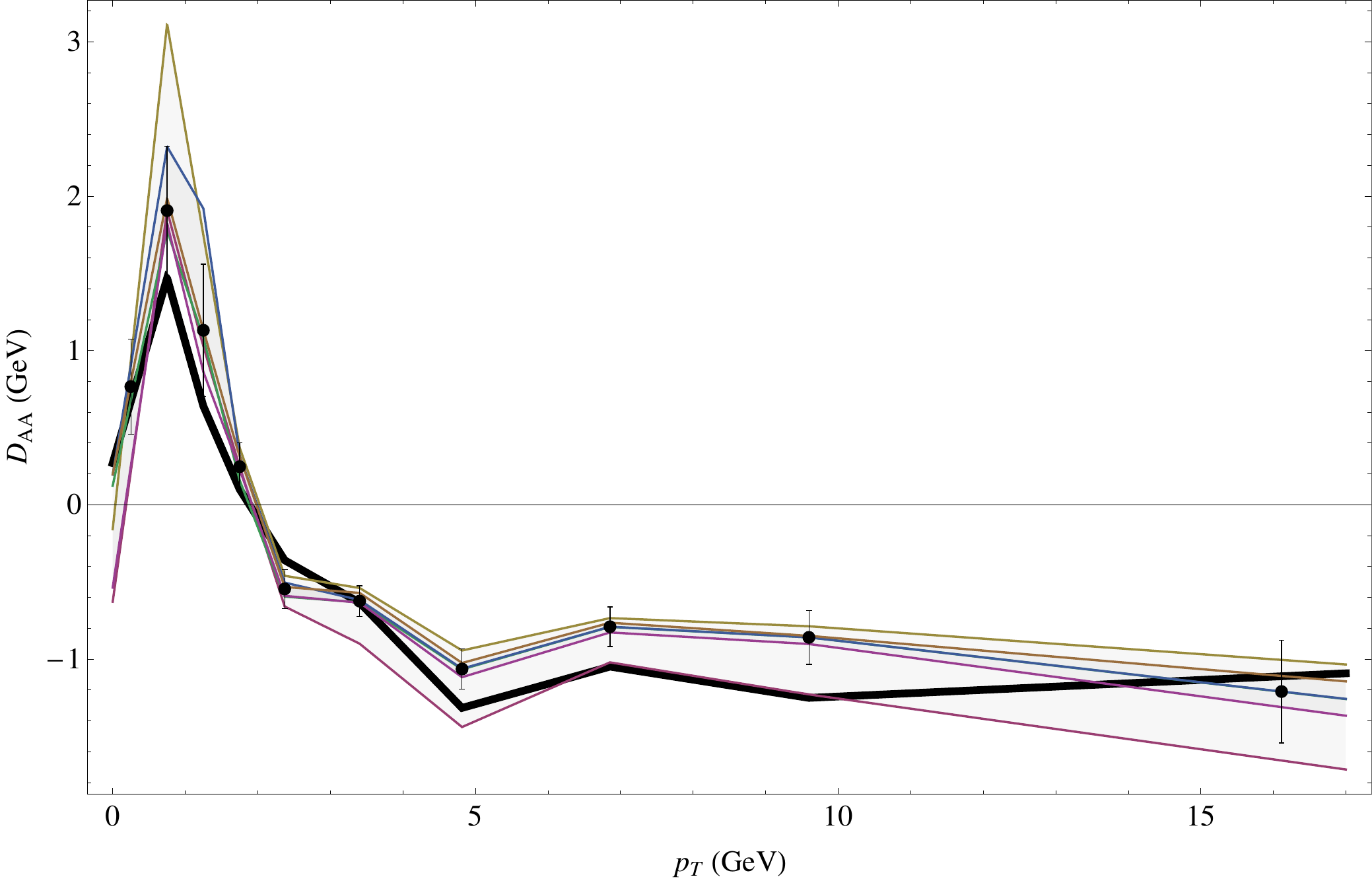} 
\caption{$D_{AA}$ defined in \eqref{eq:daa} and \eqref{eq:daa1} for jet energy of $20<p_{jet}<40$ GeV. Black dots and shaded areas show experimenal data, jet energy scale, v2/v3 and detector uncertainties respectively (taken from \cite{Adamczyk:2013up}); solid line interpolates between calculated values of $D_{AA}$ from \eqref{eq:daa1}.}
\label{fig:DAA40GeV_2}
\end{figure}

Let us note that even though the energy cannot be transferred outside the jet cone in our $1+1$ model, it is transferred from the high energy jet to low energy hadrons. Below a certain experimental cutoff, these soft hadrons are not counted as a part of the jet, and therefore this leads to an effective energy loss. To illustrate this, we compute within our model the quantity defined in \cite{Adamczyk:2013up} (in $1+1$ dimensions there is only one spatial direction, so $p_T=p$) that measures the difference in momentum distributions of the hadrons produced in get fragmentation in $AA$ and $pp$ collisions:
\begin{equation}
D_{AA}(p)=Y_{Au-Au}(p)\avg{p}_{Au-Au}-Y_{p-p}(p)\avg{p}_{p-p}
\label{eq:daa}
\end{equation}
where $Y(p)$ is the yield in a given bin with average $\avg{p}$
\begin{equation}
Y(p)=\int_{\mathrm{bin \ with \ average \ \avg{p}}}{dp'\frac{dN}{dp'}}
\label{eq:yp}
\end{equation} 
leading to
\begin{equation}
D_{AA}(p)=\avg{p}\int_{\mathrm{bin \ with \ average} \avg{p}}{dp'\frac{dN^\mathrm{med}}{dp'}}-\avg{p}\int_{\mathrm{bin \ with \ average} \ \avg{p}}{dp'\frac{dN^\mathrm{vac}}{dp'}}
\label{eq:daa1}
\end{equation}
We have plotted the results in Fig. \ref{fig:DAA40GeV_2}; we have used the mean free path of $\lambda = 0.4$ fm.
The fluctuations around $p=1$ GeV and $p=5$ GeV seem to be due to the sensitivity of $D_{AA}$ to the boundaries of the bin. The agreement with the data suggests that our simple model adequately captures the dynamics of the jet energy redistribution in the longitudinal direction. 
\vskip0.3cm 

To summarize, we have used an effective $1+1$ dimensional quasi-Abelian model to describe the dynamical modification of jet fragmentation in the QCD medium. We have found that this approach describes well the suppression of the in-medium fragmentation at intermediate values of $\xi=\ln(1/z)$ observed by the CMS Collaboration, and there is thus no contradiction between the LHC results and the picture of QCD radiation induced by the scattering of the jet.  The physics that underlies the suppression of the in-medium fragmentation function  is the partial screening of the color charge of the jet by the comoving medium-induced gluon. It would be interesting to develop a hybrid approach to jet fragmentation combining the full DGLAP perturbative evolution down to the scale of $Q_0 \sim 1 - 2$ GeV, induced gluon radiation, and the non-perturbative dynamical fragmentation as modeled above. 
\vskip0.3cm

We thank G. Milhano and J. Putschke for useful discussions. 
This work was supported in part by the U.S. Department of Energy under Contracts DE-AC02-98CH10886 and DE-FG-88ER41723.

\bibliographystyle{unsrt}

\begin{thebibliography}{99} \frenchspacing


\bibitem{CMS:2012wxa} 
  CMS Collaboration,
  CMS-PAS-HIN-12-013; \\
  F. Ma {\it et al}, [CMS Collaboration], Proc. Quark Matter 2012, {\it to appear}.


\bibitem{Gyulassy:1993hr} 
  M.~Gyulassy and X.~-n.~Wang,
  Nucl.\ Phys.\ B {\bf 420}, 583 (1994)
  [nucl-th/9306003].

\bibitem{Baier:1996kr}
  R.~Baier, Y.~L.~Dokshitzer, A.~H.~Mueller, S.~Peigne, D.~Schiff,
  Nucl.\ Phys.\  {\bf B483}, 291-320 (1997).
  [hep-ph/9607355]; Nucl.\ Phys.\  {\bf B484}, 265-282 (1997).

\bibitem{Gyulassy:2000er}
  M.~Gyulassy, P.~Levai, I.~Vitev,
  Nucl.\ Phys.\  {\bf B594}, 371-419 (2001).
  [nucl-th/0006010].

\bibitem{CasalderreySolana:2010eh}
  J.~Casalderrey-Solana, J.~G.~Milhano, U.~A.~Wiedemann,
  J.\ Phys.\ G {\bf G38}, 035006 (2011).

\bibitem{Zakharov:1996fv}
  B.~G.~Zakharov,
  JETP Lett.\  {\bf 63}, 952 (1996)
  [hep-ph/9607440].

\bibitem{Baier:2000mf}
  R.~Baier, D.~Schiff and B.~G.~Zakharov,
  Ann.\ Rev.\ Nucl.\ Part.\ Sci.\  {\bf 50}, 37 (2000)
  [hep-ph/0002198].

\bibitem{Armesto:2011ht} 
  N.~Armesto, B.~Cole, C.~Gale, W.~A.~Horowitz, P.~Jacobs, S.~Jeon, M.~van Leeuwen and A.~Majumder {\it et al.},
  arXiv:1106.1106 [hep-ph].

\bibitem{Dokshitzer:1991wu}
  Y.~L.~Dokshitzer, V.~A.~Khoze, A.~H.~Mueller, S.~I.~Troian,
  ``Basics of perturbative QCD,''
  Gif-sur-Yvette, France: Ed. Frontieres (1991) 274 p. 


\bibitem{Loshaj:2011jx} 
  F.~Loshaj and D.~E.~Kharzeev,
  Int.\ J.\ Mod.\ Phys.\ E {\bf 21}, 1250088 (2012)
  [arXiv:1111.0493 [hep-ph]].
  
\bibitem{CasalderreySolana:2012ef} 
  J.~Casalderrey-Solana, Y.~Mehtar-Tani, C.~A.~Salgado and K.~Tywoniuk,
  arXiv:1210.7765 [hep-ph].

\bibitem{Collins:1989gx} 
  J.~C.~Collins, D.~E.~Soper and G.~F.~Sterman,
  Adv.\ Ser.\ Direct.\ High Energy Phys.\  {\bf 5}, 1 (1988)
  [hep-ph/0409313].

\bibitem{Gribov:1999ui}
  V.~N.~Gribov,
  Eur.\ Phys.\ J.\  {\bf C10}, 91-105 (1999).

\bibitem{Dokshitzer:2004ie}
  Y.~L.~Dokshitzer, D.~E.~Kharzeev,
  Ann.\ Rev.\ Nucl.\ Part.\ Sci.\  {\bf 54}, 487-524 (2004).
  [hep-ph/0404216]. 

\bibitem{Schwinger:1962tp}
  J.~S.~Schwinger,
  Phys.\ Rev.\  {\bf 128}, 2425-2429 (1962).

\bibitem{Lowenstein:1971fc}
  J.~H.~Lowenstein, J.~A.~Swieca,
  Annals Phys.\  {\bf 68}, 172-195 (1971).
  
\bibitem{Coleman:1975pw}
  S.~R.~Coleman, R.~Jackiw, L.~Susskind,
  Annals Phys.\  {\bf 93}, 267 (1975).

\bibitem{:1974cks}
  A.~Casher and J. ~Kogut and L.~Susskind,
  Phys.\ Rev.\ D\  {\bf 10}, 732 (1974).

\bibitem{Fujita:1989vv}
  T.~Fujita, J.~Hufner,
  Phys.\ Rev.\  {\bf D40}, 604-606 (1989).

\bibitem{Wong:1991ub}
  C.~Y.~Wong, R.~- C.~Wang, C.~C.~Shih,
  Phys.\ Rev.\  {\bf D44}, 257-262 (1991); 
  J.~Liu, C.~Y.~Wong, C.~C.~Shih, R.~-C.~Wang,
  Phys.\ Lett.\  {\bf B326}, 154-160 (1994).


\bibitem{'tHooft:1977hy}
  G.~'t Hooft,
  Nucl.\ Phys.\  {\bf B138}, 1 (1978).
    
\bibitem{Mandelstam:1974pi}
  S.~Mandelstam,
  Phys.\ Rept.\  {\bf 23}, 245-249 (1976).    
  
\bibitem{Adamczyk:2013up} 
  L.~Adamczyk {\it et al.}  [STAR Collaboration],
  arXiv:1301.6633 [nucl-ex].
\end{thebibliography}

\end{document}